\documentclass[journal=jacsat,manuscript=article]{achemso}

\usepackage{chemformula} 
\usepackage[T1]{fontenc} 
\usepackage{upgreek}
\usepackage{xcolor}

\author{Riande I. Dekker}
\email{r.i.dekker@uva.nl}
\affiliation{Van der Waals-Zeeman Institute, Institute of Physics, University of Amsterdam, Science Park 904, 1098 XH Amsterdam, The Netherlands}
\alsoaffiliation{Van 't Hoff Laboratory of Physical and Colloid Chemistry, Debye Institute for Nanomaterials Science, Utrecht University, Padualaan 8, 3584 CH, Utrecht, The Netherlands}
\author{A. Deblais}
\affiliation{Van der Waals-Zeeman Institute, Institute of Physics, University of Amsterdam, Science Park 904, 1098 XH Amsterdam, The Netherlands}
\alsoaffiliation{Unilever Foods Innovation Center, Bronland 14, 6708 WH Wageningen, The Netherlands}
\author{Krassimir P. Velikov}
\affiliation[UVA]{Van der Waals-Zeeman Institute, Institute of Physics, University of Amsterdam, Science Park 904, 1098 XH Amsterdam, The Netherlands.}
\alsoaffiliation{Unilever Foods Innovation Center, Bronland 14, 6708 WH Wageningen, The Netherlands}
\alsoaffiliation{Soft Condensed Matter, Debye Institute for Nanomaterials Science, Utrecht University, Princetonplein 5, 3584 CC, Utrecht, The Netherlands}
\author{Peter Veenstra}
\affiliation{Shell Global Solutions International B.V., Grasweg 31, 1031 HW Amsterdam, The Netherlands}
\author{Annie Colin}
\affiliation{Chimie Biologie Innovation, ESPCI Paris, CNRS, PSL University, 10 rue Vauquelin, 75005 Paris, France}
\alsoaffiliation{Centre de Recherche Paul Pascal, CNRS, Universit\'e de Bordeaux, 115 Avenue Schweitzer, 33600 Pessac, France}
\author{Hamid Kellay}
\affiliation{Laboratoire Ondes et Matie\`ere d'Aquitaine, UMR 5798, CNRS, Universit\'e de Bordeaux, 351 Cours de la Lib\'eration, 33400 Talence, France}
\author{Willem K. Kegel}
\affiliation{Van 't Hoff Laboratory of Physical and Colloid Chemistry, Debye Institute for Nanomaterials Science, Utrecht University, Padualaan 8, 3584 CH, Utrecht, The Netherlands}
\author{Daniel Bonn}
\affiliation{Van der Waals-Zeeman Institute, Institute of Physics, University of Amsterdam, Science Park 904, 1098 XH Amsterdam, The Netherlands}

\title{Emulsion Destabilisation by Squeeze Flow}

\abbreviations{IR,NMR,UV}
\keywords{American Chemical Society, \LaTeX}

\begin{document}


\newpage

\begin{abstract}
There is a large debate on the destabilisation mechanism of emulsions. We present a simple technique using mechanical compression to destabilise oil-in-water emulsions. Upon compression of the emulsion, the continuous aqueous phase is squeezed out, while the dispersed oil phase progressively deforms from circular to honeycomb-like shapes. The films that separate the oil droplets are observed to thin and break at a critical oil/water ratio, leading to coalescence events. Electrostatic interactions and local droplet rearrangements do not determine film rupture. Instead, the destabilisation occurs like an avalanche propagating through the system, starting at areas where the film thickness is smallest.
\end{abstract}

\section{Introduction}
Destabilisation of an emulsion or foam occurs when individual droplets that make up the system start to coalesce, breaking the film of liquid in between them. The mechanism of destabilisation of foams and emulsions is still a matter of considerable debate, in spite of the fact that it is of paramount importance for many processes and applications. Much attention has focused on the mechanism behind the destabilisation in foams. Some studies report that foam coalescence is induced when the capillary pressure exceeds the disjoining pressure \cite{Khristov2002,Feng2013}, making the colloidal interactions determine the stability. Other studies however suggest that the interactions are not important and that local rearrangements are at the origin of the destabilisation: the drainage leads to a shortage of liquid for making transient films during rearrangements and hence the films break when attempting such a rearrangement \cite{Carrier2003, Biance2011}. More recently, Forel \textit{et al.} \cite{Forel2019} claim, in contradiction with both mechanisms, that foam destabilisation is due to film rupture which is simply a stochastic process with the probability of rupture being proportional to the film area. The discussion of the foam stability is complicated by the large number of effects that are present; gravity, disjoining pressure, geometry etc., making it difficult to decide between the different scenarios of destabilisation.

Emulsions are very similar to foams, and their stability similarly important for applications \cite{Bibette1992,Wasan2004}. For instance, the destabilisation of emulsions is a key step in oil recovery, to extract water from the recovered crude oil  \cite{White2000,Pena2005,Less2008,Katepalli2016}. A key advantage of emulsions over foams is that gravity is much less important because the density of water and oil phases are similar. Furthermore, the two phases can be refractive-index matched, so that their (in)stability can be investigated relatively easily using scanning confocal microscopy. 

In this article we exploit these advantages to investigate the mechanism of emulsion destabilisation in a highly concentrated oil-in-water emulsion. Through mechanical compression of an emulsion, we induce syneresis as water is squeezed out of the system. Using confocal microscopy we follow the behaviour of individual droplets, looking for the role of the water film in between droplets and of colloidal interactions in the process of destabilisation. Furthermore, we see that this new technique allows for easy investigation of various trends in the destabilisation process needing only small volumes.

\begin{figure}
	\includegraphics[width=16cm]{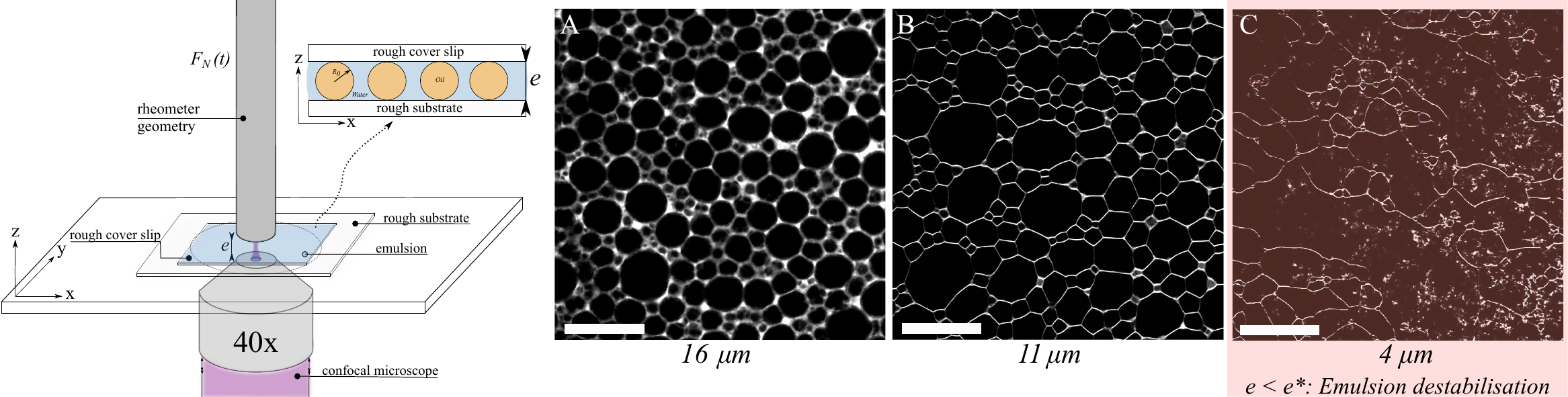}
	\caption{\textbf{Emulsion destabilisation experiment}. Left: schematic picture of the experimental setup (not to scale): an initial volume $V_{0}$ of emulsion is squeezed between two glass plates which consist of a thick glass plate (1~mm) and a thin (170~$\upmu$m) microscope cover glass slide on top. A position-controllable rheometer head allows to impose the desired thickness $e$ to the emulsion layer. This setup is mounted on top of a confocal microscope stage. Right: consecutive confocal images of a typically squeezing experiment, showing deformation of the emulsion (the continuous phase is rendered bright) under compression. Strong deformation of the emulsion occurs until a critical thickness $e^*$ is reached, after which partial and eventually total destabilisation of the emulsion is observed. The scale bars correspond to a distance of 50~$\upmu$m.} 
	\label{emulsion_destabilisation}
\end{figure}

\section{Results and discussion}
The measurement setup and typical images of our destabilisation experiments are shown in Fig.~\ref{emulsion_destabilisation}. We use an 80 v\% silicone-oil-in-water emulsion stabilised by 1 wt\% sodium dodecyl sulfate (SDS) with an average droplet size of 18 $\upmu$m and a dispersity of 20\%. Advantages of using a polydisperse system is that it is prevented from possible crystallisation that can easily happen in a monodisperse emulsion under pressure \cite{Bibette1992} and that the system resembles everyday life emulsions, such as mayonnaise \cite{McClements2015}. An initial volume $V_0$ of the emulsion is squeezed between a rough glass plate of 1 mm thickness and a thin microscope cover glass slide. The rough bottom plate prevents the emulsion from sliding and keeps the oil droplets in position \cite{Bertola2003,Paredes2015}. In this way, only the continuous phase is evacuated leading to a gradual change of the volume fraction of the emulsion. However, the droplets are not pinned and can still undergo rearrangements, as will be visualised later in this paper. 

A position-controllable rheometer head is mounted on top of a confocal microscope stage and imposes the desired thickness $e$ to the emulsion layer. Our technique provides unprecedented control over the internal phase volume fraction while simultaneously providing a direct visualisation of the emulsion structure. This allows us to measure the surface coverage from our confocal images. As we can alter the thickness of the emulsion is small steps of about 1 $\upmu$m, the internal phase surface coverage only increases by a few percent per step. Therefore, we believe that we know the critical internal phase surface coverage with only 1 to 2 $\%$ uncertainty. During the measurements, the imaging settings are kept constant.

The confocal microscopy images (Fig.~\ref{emulsion_destabilisation}, Panels A-C) of different sample thicknesses first reveal a transition from nearly spherical to polygonal oil droplets (black) in the continuous phase (white) during compression, similar to the observations of Morse \textit{et al.} \cite{Morse1993}. Overall, the emulsion is only slightly confined and the emulsion remains stable. The polygonal shape, which is reminiscent of a honeycomb structure, has also been observed previously, both in emulsions and in liquid foams \cite{Princen1979,Princen1980,Drenckhan2015}. Upon decreasing the sample thickness even further, we observe large facets of oil droplets pressed against each other, with a very thin film of water in between. Once a certain critical thickness $e^*$ is reached, coalescence events between oil droplets start to occur and the emulsion becomes unstable. Confining the droplets even further leads to more coalescence events and eventually full destabilisation of the emulsion (Fig.~\ref{emulsion_destabilisation}, Panel C). This transition can be quantified in terms of the surface fraction of oil, which  increases with decreasing sample thickness to over 90 \% at the critical sample thickness $e^{*}$. The water films between the oil droplets also become thinner with decreasing sample thickness, indicating that most of the continuous phase is being squeezed out from the emulsion layer. Complete 3D-images of the emulsion during the squeezing experiments are recorded, but careful investigation in the z-direction does not lead us to think that 3D-effects are important in our system.

One of the reported reasons for foam destabilisation is an imbalance between the capillary pressure and the disjoining pressure \cite{Khristov2002}. Of course, as the sample thickness decreases, the surfaces between the droplets become flatter. However, as already observed by Morse \textit{et al.} \cite{Morse1993}, the contact area of interaction remains slightly curved and in a stable emulsion the disjoing pressure is equal to the Laplace pressure. The capillary pressure can be described by the Young-Laplace equation $\Delta p = \gamma / 2R$ where $\Delta p$ is the capillary pressure, $\gamma$ the oil-water surface tension, and $R$ the droplet radius. As the emulsion consists of soft, compressible droplets, the droplets deform during the squeeze experiments. This results in a change in the capillary pressure as $R$ changes. The disjoining pressure consists of a Van der Waals component and an electrostatic component. This electrostatic component is dependent on the ionic strength of the continuous phase via the Debye length $\Pi_{el} = C \exp(-\kappa d)$ where $\Pi_{el}$ is the electrostatic component of the disjoining pressure, $C$ a constant depending on the ionic concentration $c$, temperature, electronic charge and surface potential, $\kappa$ the reciprocal Debye length ($\kappa \propto \sqrt{c}$) and $d$ the film thickness \cite{Binks1997}. Since the disjoining pressure is clearly dependent on the ionic concentration, the droplet size $R$ where the capillary pressure no longer balances the disjoining pressure should therefore also be dependent on the ionic strength. 

To investigate the role of the electrostatic interactions in the destabilisation process, we compare three emulsions with different salt concentrations in the continuous phase. The dependence of emulsion destabilisation on the salt concentration has been reported in literature \cite{Neumann2004,Krebs2013}. The type of measurements to determine the stability of emulsions in these papers deviates from our experimental setup. Neumann \textit{et al.} \cite{Neumann2004} used a coalescence cell in which droplets were formed in an aqueous phase. The droplets were classified as stable when their lifetimes exceeded 30 minutes. Krebs \textit{et al.} \cite{Krebs2013} studied the stability of emulsions with low surfactant concentration upon shear-induced collisions using a microfluidics device. In our system, the surfactant concentration remains well above the critical micelle concentration and variations in the salt concentration in the continuous phase are applied, thereby controlling the osmotic pressure of the system. We use the 1:1 electrolyte sodium chloride. When no additional NaCl is added, the presence of 1 wt\% of SDS already leads to an ionic strength of 35 mM, resulting in $\kappa^{-1} = 1.63$ nm. Addition of 10 mM of NaCl results in $\kappa^{-1} = 1.44$ nm and 25 mM of additional NaCl results in $\kappa^{-1} = 1.24$ nm. Large differences in the Debye screening length are not achieved, due to limitations in the solubility of SDS in a saline solution.   

\begin{figure}
	\includegraphics[width=16cm]{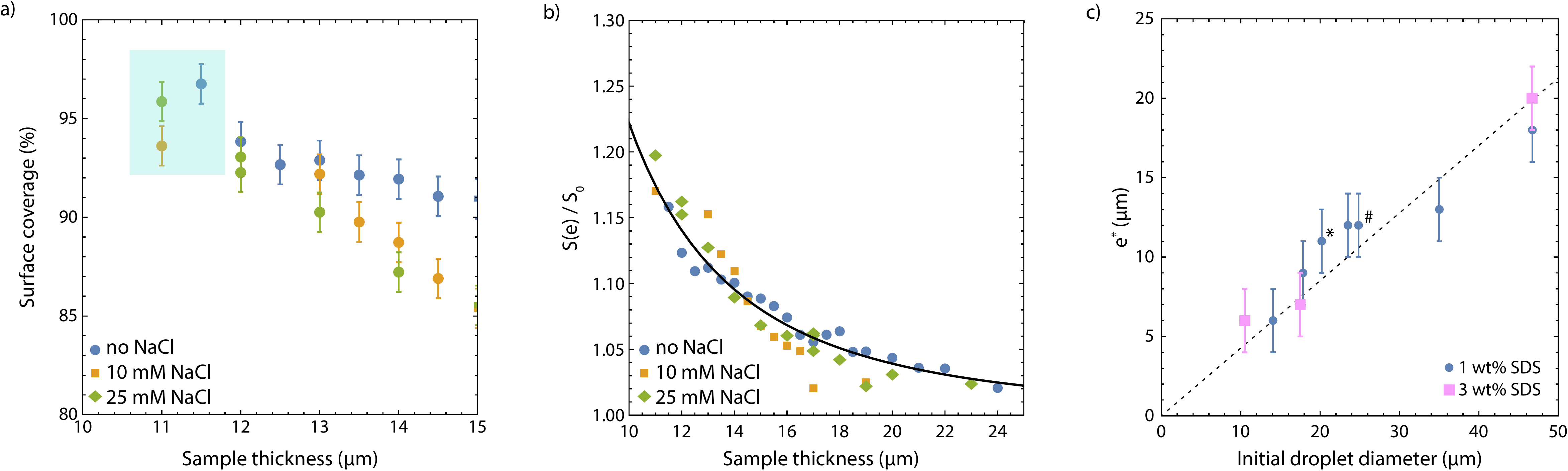}
	\caption{\textbf{Effect of the ionic strength on the critical sample thickness $e^*$ for film rupture}. (a) Surface coverage $S$ as a function of sample thickness for different salt concentrations. The light blue square shows the region where emulsion destabilisation starts. (b) Surface coverage $S$ (rescaled to its initial value $S_0$) as a function of sample thickness for different salt concentrations. The solid line is a guide to the eye: $S(e)/S_{0} = 1 + b e^{m}$ with $m$ = -2.5 and $b$ = 70~$\upmu$m$^{5/2}$.  (c) Critical sample thickness at which coalescence occurs as a function of the initial droplet diameter $d_i$ for two different surfactant (SDS) concentrations. The * symbol highlights the data point with 10 mM of NaCl added to the continuous phase and the \# symbol shows the sample with 25 mM of NaCl. The dashed line is a linear fit of the data: $e^{*} \simeq$ 0.43 $d_{i}$.} 
	\label{saltconcentration}
\end{figure}

Fig.~\ref{saltconcentration}(a) shows the oil surface coverage as a function of the thickness of the emulsion layer during the squeezing experiments for three emulsions with different salt concentrations. The light blue square highlights the region of emulsion break up. In Fig.~\ref{saltconcentration}(b) the surface coverage of oil is scaled by the initial surface coverage $S_{0}$. $S_{0}$ is the surface coverage measured at relatively large sample thickness, to make sure that the droplets are not yet deformed. Although all prepared emulsions have a volume fraction of oil of 0.8, the surface fraction of oil is always slightly lower and small variations are observed between the various samples. We observe the same increase of oil surface coverage with decreasing sample thickness for all salt concentrations with a critical sample thickness $e^{*}$ of around 11~$\upmu$m where the first coalescence events start to occur. At the critical sample thickness, the surface fraction of all samples has increased to approximately 95 \%. 

To further investigate the trend in emulsion destabilisation, emulsions with different initial droplet size are prepared. The initial droplet size is dependent on the rotation speed during mixing: mixing speeds between 2 and 8 krpm are used, resulting in droplets between 14 and 47 $\upmu$m. The emulsions all have a dispersity of around 20$\%$. Besides addition of NaCl to change the ionic strength of the continuous phase, various emulsions with a higher concentration of 3 wt\% surfactant are investigated, which also increases the ionic strength. Furthermore, as the surfactant provides the stability of the emulsion, a higher surfactant concentration is likely to influence the destabilisation. Feng \textit{et al.} \cite{Feng2013} showed that coalescence in a drying emulsion was slowed down by increasing the SDS concentrations.

As shown in Fig.~\ref{saltconcentration}(c), the critical thickness for coalescence is an approximately linear function of the initial droplet size. Altering the strength of electrostatic interactions by changing the surfactant (purple squares in Fig.~\ref{saltconcentration}(c)) or salt concentration (see data points highlighted with * and \#) in the continuous phase does not influence this trend. These results indicate that film rupture is not due to an imbalance between capillary and disjoining pressure, and that the stability is thus not determined by colloidal interactions. However, we do observe a higher critical surface fraction when using 3 wt\% of SDS instead of 1 wt\% SDS. This higher surface fraction that needs to be reached for film rupture does not influence the critical sample thickness, though, it can explain the slowing down of coalescence in a drying emulsion \cite{Feng2013}. We do note that the range of ionic strenghts in our research is limited and think that it would be interesting to investigate the effect of very high ionic strength of the emulsion destabilisation.

\begin{figure}
	\includegraphics[width=16cm]{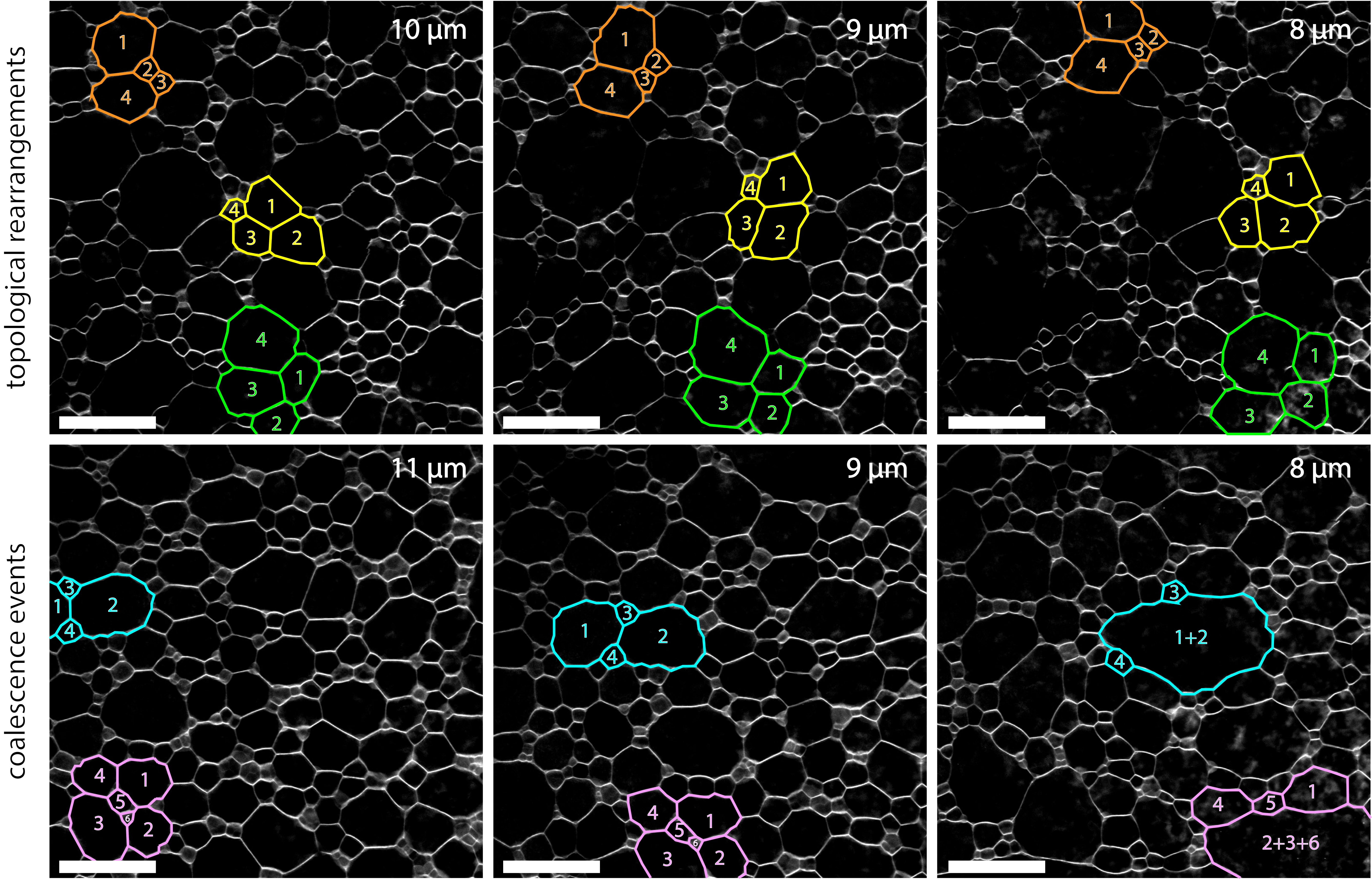}
	\caption{
		\textbf{Local droplet arrangements.} Confocal image series of the emulsion during a squeeze experiment. The sample thickness in both cases is around the critical sample thickness of 10~$\upmu$m. Upper row: three subsequent images showing multiple local rearrangements. Bottom row: three subsequent images showing film rupture resulting in coalescence of two droplets. Individual droplets are marked using color outlines and numbers. The sample thickness is reduced quasi-statically by approximately 1~$\upmu$m at a time. There are about 6 minutes between each of the consecutive images. The scale bars correspond to a distance of 50~$\upmu$m.} 
	\label{T1rearrangements}
\end{figure}

The increase in oil surface coverage before film rupture resembles observations in foams by Carrier and Colin \cite{Carrier2003} of a critical liquid fraction. They mention that below a critical liquid fraction rearrangements lead to rupture of thin films between foam bubbles. These so-called T1 rearrangements are very rapid movements of four droplets changing their position relative to each other \cite{Carrier2003,Biance2011}. According to these observations for the coalescence in foams, T1 rearrangements would only be observed above a critical sample thickness when the liquid fraction is still high enough. Below the critical sample thickness, the liquid fraction would be too low for T1 rearrangements to occur. The upper sequence of confocal images in Fig.~\ref{T1rearrangements} shows an example of these local rearrangements in an emulsion layer with a thickness around the critical thickness for film rupture. Here, the emulsion remains intact despite the presence of rearrangements. The bottom sequence of confocal images highlights two events of film rupture, thereby inducing emulsion destabilisation. The sample thickness is similar for both sets of confocal images. To further investigate the presence of T1 rearrangements and film rupture, we carefully examine fifteen experiments. In all these experiments, we make sure to go well beyond the critical sample thickness for coalescence. Two experiments do not show any rearrangements or film rupture events, caused by a too rapid decrease in the sample thickness. Five experiments only show film rupture, without the presence of rearrangements and four experiments show rearrangements event below the critical sample thickness without clear film rupture events. It is good to mention that for these four experiments we have clear proof that coalescence occurs, although no individual film rupture events can be observed. Four experiments show both rearrangements and film rupture events below the critical sample thickness, but no rearrangements that lead to film rupture. Rearrangements always occur at regions free from film rupture. Generically,  the results show that around the critical thickness for destabilisation, local rearrangements are still possible but do not induce coalescence. Rather, coalescence occurs in regions devoid of rearrangements. We conclude that film rupture occurs independently of any rearrangements. 

\begin{figure}
	\includegraphics[width=16cm]{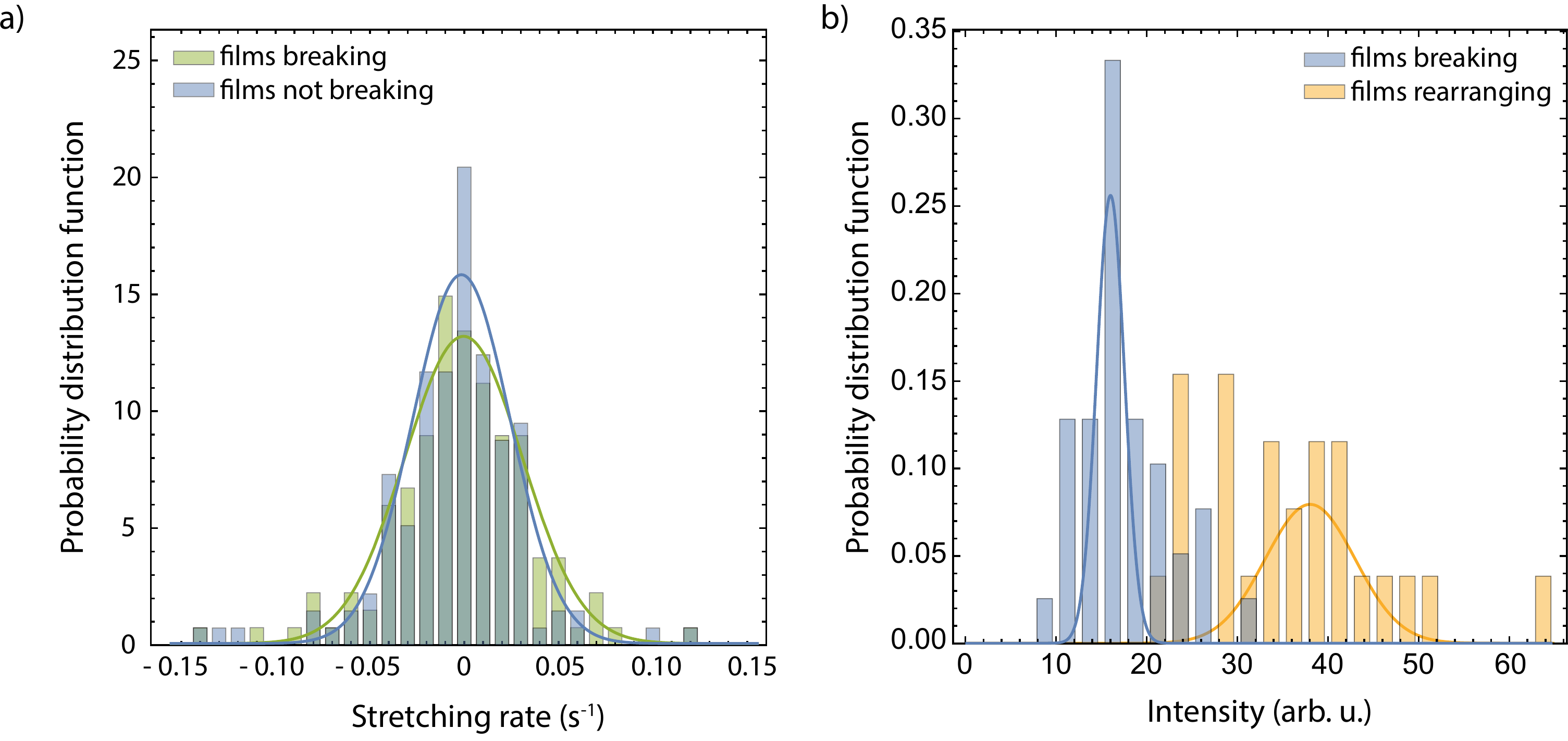}
	\caption{\textbf{Rupturing versus non-rupturing films.} (a) Normalised probability distribution functions of stretching rates for rupturing (green) and non-rupturing (blue) films. Lines are Gaussian fits. (b) Probability distribution functions of the confocal light intensity of films shortly before rupture (blue) compared to films not rupturing but undergoing a rearrangement (yellow). Lines are Gaussian fits.} 
	\label{stretchingrate}
\end{figure}

The question remains what causes certain films to break. We therefore examine characteristic features of rupturing films and compare them to those of films that do not rupture but undergo a rearrangement. We expect the film to shrink in the z-dimension, as we decrease the distance between the slides. We observe that the effect of shrinking in the z-dimension does affect the length in the lateral direction, but this stretching fluctuates from film to film. Measurements of stretching rates for breaking and non-breaking films, carried out for a total time of one minute, confirm the presence of such stretching. The stretching rate is defined as $\frac{1}{L} \frac{dL}{dt}$ with $L$ the length of the film in between droplets and $dL$ the change in length over a period $dt$ of one second. The normalised probability distribution functions (PDFs) of the stretching rates are shown in Fig.~\ref{stretchingrate}(a). The PDFs for the breaking and non-breaking films seem to overlap for the largest part, however small variations can be observed. The PDFs are fit with a Gaussian function

\begin{equation}
f(x) = \frac{1}{\sigma \sqrt{2\pi}}  e^{\frac{-(x-\mu)^2}{2 \sigma^2}},
\label{Gaussian}
\end{equation}

where $\sigma^2$ is the variance and $\mu$ is the expected value, which is around 0. For the films that do not break a slightly higher probability around 0 is observed, indicating that these non-breaking films are subjected to less stretching. The breaking films show slightly higher probabilities at both positive and negative stretching rates as well as a higher variance ($\sigma$ = 0.030 for breaking films and $\sigma$ = 0.025 for non-breaking films). These measurements show that fluctuations in stretching rate are present during the squeeze experiments for breaking and non-breaking films.

We also investigate differences in the film thickness in rupturing versus non-rupturing films. A measure of this thickness is the light intensity as recorded in a confocal image. Fig.~\ref{stretchingrate}(b) shows the PDFs of the intensities of films seconds before breaking versus undergoing a rearrangement over an image width of 4~$\upmu$m. Two separate peaks can clearly be distinguished. The films that rupture have a much lower intensity than the films undergoing a rearrangement, meaning that breaking films are much thinner compared to films involved in a rearrangement. From these differences in intensities between breaking and rearranging films, we conclude that thin films have a higher probability of rupture, whereas thicker films can still undergo rearrangements without inducing film rupture. 

\begin{figure}
	\includegraphics[width=16cm]{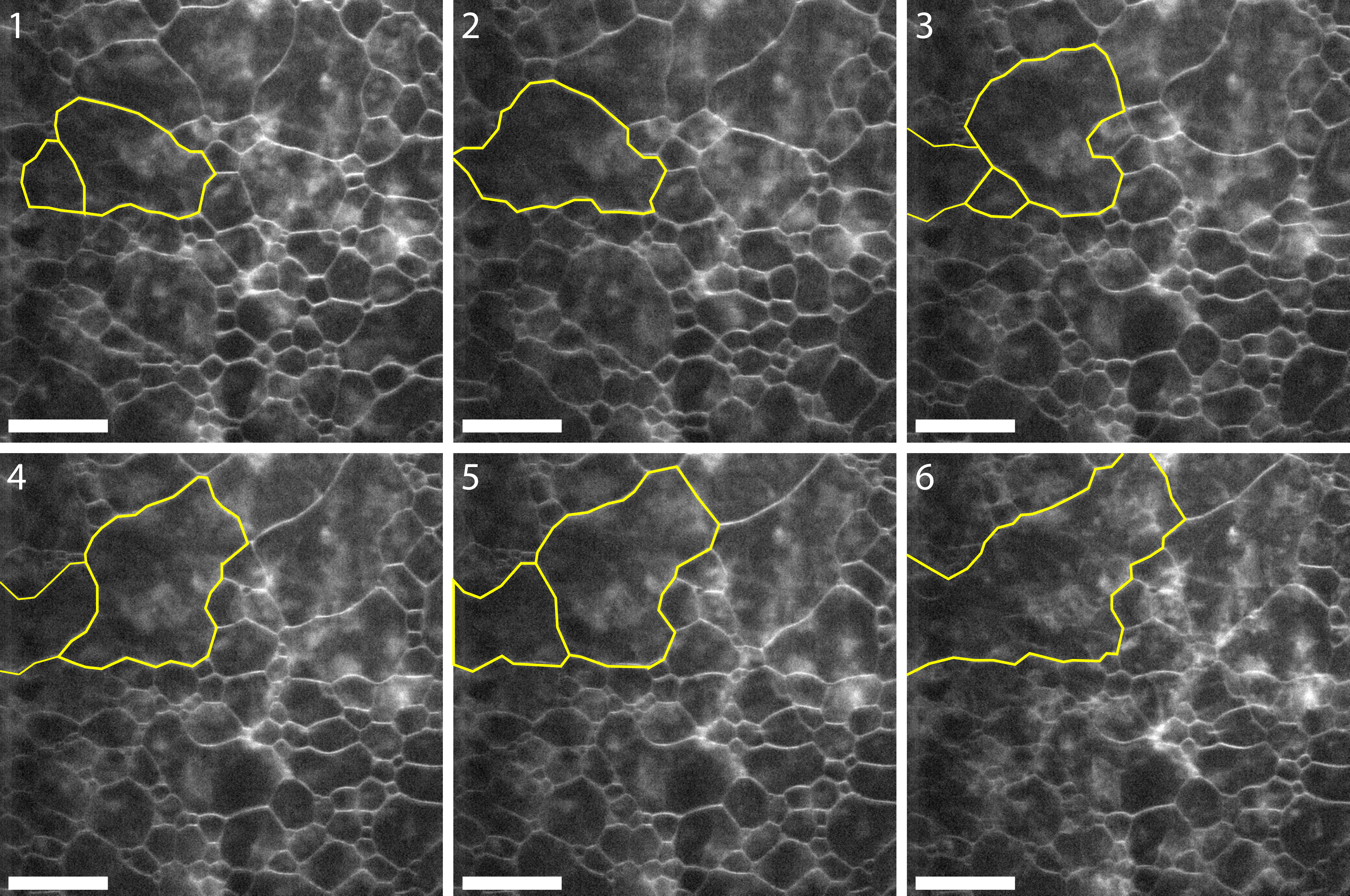}
	\caption{
		\textbf{Cascade  of  coalescence  events.} Consecutive confocal images recording with 1 - 2 minute intervals. Coalescing droplets are highlighted in yellow. The scale bars correspond to a distance of 50~$\upmu$m.} 
	\label{cascade}
\end{figure}

We observe that the first film rupture causes the surrounding droplets to rearrange. However, as the films separating the droplets have already become very thin, this leads to coalescence events instead of regular rearrangements. Fig.~\ref{cascade} shows a cascade of coalescence events induced by a first break up of a very thin film. The rest of the emulsion layer remains stable and hardly moves during these coalescence events. The first film rupture seems to be a stochastic process depending on the thickness, and resembles the results by Forel \textit{et al.} \cite{Forel2019}. However, this first coalescence event induces further coalescence events which are not simply stochastic anymore. For a more thorough investigation of the cascade of coalescence events we perform ten experiments at a sample thickness slightly below the critical sample thickness for coalescence. At this point the first film rupture events have already occured and from the results in Fig.~\ref{cascade} we thus expect a cascade of coalescence events finally resulting in very large droplets and complete destabilisation of the emulsion. Only one out of the ten experiments does not show this cascade, which is due to large movements of the emulsion as a whole, because the pinning of the emulsion droplets by the rough substrate is not sufficient in this case. The other nine experiments clearly show coalescence cascades, in five of them multiple cascade events occur simultaneously caused by two or three film ruptures at different spots in the emulsion at similar times. As the films between the droplets become thinner everywhere in the emulsion, it is well understandable that multiple film ruptures can occur at the same time.

In this paper, we investigated only a limited range of droplet sizes, volume fractions and salinities. Nevertheless, we believe that our findings apply to a broad range of concentrated emulsions. Emulsions with initial droplet sizes between 10 and 50~$\upmu$m are investigated, but preliminary experiments on emulsions with oil droplets of only a few micrometer give similar results. Below 1~$\upmu$m, thermal stresses start to play a role, which might strongly influence the destabilisation process. We report the results of emulsions with 80 v$\%$ of oil, however, we have performed measurements on volume fractions ranging from 70 - 80 v$\%$ of oil. From these experiments, we see that the critical volume fraction is independent of the initial volume fraction for emulsions above the critical volume fraction for jamming. At low volume fractions of oil, the oil droplets are not jammed and we therefore expect different behaviour upon squeezing, where the oil droplets might as well squeeze out. The limited range of salinities that we have investigated is discussed above. Within the limits of our investigations, we have observed that the destabilisation mechanism is purely geometrical. We conclude that neither the disjoining pressure nor the capillary force plays a role. Therefore, the reduction in sample thickness to achieve destabilisation of the emulsion can be predicted from the initial volume fraction and average droplet size. We find that destabilisation starts at a surface coverage of roughly 95 v$\%$. This geometric mechanism behind destabilisation is very generic and can be used for similar jammed systems.

\section{Conclusions}
We have described a simple technique to destabilise surfactant-stabilised oil-in-water emulsions. It leads to an increase in the oil fraction by preferentially squeezing out water, resulting in a critical oil fraction where the thinnest films rupture first. We compare our destabilising emulsion with literature about foam destabilisation and show that neither electrostatic interactions nor local rearrangements explain our destabilisation mechanism. We find that the probability for the first rupture event increases with increasing oil fraction as this results in thinner films that cannot not resist stretching. A cascade of coalescence events can then be observed as the result of film movement triggered by the first film rupture. Furthermore, this new technique can be of great interest in, for example, the oil industry. Production chemicals that are used to extract oil can have large impact on the stability of the emulsion. The squeeze flow experiments allow to investigate these effects on a small scale. Interesting would be to shift towards more realistic conditions, e.g. by measuring in the presence of natural gas.

\begin{acknowledgement}
	
This work is part of the research programme Controlling Multiphase Flow with project number 680-91-012, which is (partly) financed by the Dutch Research Council (NWO) and co-funded by TKI-E\&I with the supplementary grant 'TKI- Toeslag' for Topconsortia for Knowledge and Innovation (TKI’s) of the Ministry of Economic Affairs and Climate Policy. This work took place within the framework of the Institute of Sustainable Process Technology. The authors thank the workshop of the University of Amsterdam for their technical assistance. A.D acknowledges funding from the European Union’s Horizon 2020 research and innovation programme under the Individual Marie Skłodowska-Curie fellowship grant agreement number 798455. This work was partially funded by Evodos, Shell and Unilever R\&D.
	
\end{acknowledgement}

\providecommand{\latin}[1]{#1}
\makeatletter
\providecommand{\doi}
  {\begingroup\let\do\@makeother\dospecials
  \catcode`\{=1 \catcode`\}=2 \doi@aux}
\providecommand{\doi@aux}[1]{\endgroup\texttt{#1}}
\makeatother
\providecommand*\mcitethebibliography{\thebibliography}
\csname @ifundefined\endcsname{endmcitethebibliography}
  {\let\endmcitethebibliography\endthebibliography}{}

\end{document}